\documentclass[twocolumn,preprintnumbers,aps]{revtex4}
\usepackage{graphicx}
\usepackage{dcolumn}
\usepackage{bm}

\begin{document}

\title{Electronic Structure and Optical Properties of \\
Silicon Nanocrystals along their Aggregation Stages}

\author{C. Bulutay}
\email{bulutay@fen.bilkent.edu.tr}
\affiliation{Department of Physics, Bilkent University, Bilkent,
 Ankara, 06800, Turkey}
\date{\today}

\begin{abstract}
The structural control of silicon nanocrystals is an important technological problem. 
Typically a distribution of nanocrystal sizes and shapes emerges under the uncontrolled 
aggregation of smaller clusters. The aim of this computational study is to investigate 
the evolution of the nanocrystal electronic states and their optical properties 
throughout their aggregation stages. To realistically tackle such systems, an atomistic 
electronic structure tool is required that can accommodate 
about tens of thousand nanocrystal and embedding lattice atoms with very irregular shapes. 
For this purpose, a computationally-efficient pseudopotential-based electronic structure tool 
is developed that can handle realistic nanostructures based on the expansion of the wavefunction 
of the aggregate in terms of bulk Bloch bands of the constituent semiconductors. With this tool, 
the evolution of the electronic states as well as the polarization-dependent absorption spectra 
correlated with the oscillator strengths over their aggregation stages are traced. 
The low-lying aggregate nanocrystal states develop  binding and anti-binding counterparts of the 
isolated states. Such information may become instrumental with the maturity of the controlled 
aggregation of these nanocrystals.
\end{abstract}

\maketitle
\section{Introduction}
Embedded semiconductor nanocrystals (NCs) can be fabricated by several different techniques such as ion
implantation~\cite{iimp1,iimp2}, sputtering~\cite{sput}, plasma enhanced chemical vapor 
deposition~\cite{pecvd}. However, their structural control is still an important technological problem. 
During the annealing process, the excess silicon atoms start forming clusters and with their gradual 
aggregation into nanocrystals, a distribution of sizes and shapes emerges.
The annealing temperature and duration determine the degree of the cluster aggregation and eventually 
the average NC size distribution. Being directly controlled by this distribution~\cite{pecvd,brongersma} 
the luminescence properties are intimately affected by this process of NC aggregation driven 
by atomistic nucleation and growth kinetics~\cite{heinig}. It has been generally demonstrated that 
the increase in the Si concentration of the substoichiometric silicon oxide films, as well as the 
annealing temperature lead to larger NCs and hence to the red shift of their luminescence 
spectra~\cite{pecvd}. However, the experimental studies are hampered by the large size distribution 
existing in typical samples. The soul of the present work is based on the controlled agglomeration of 
Si NCs. This is currently not achievable by the PECVD, sputtering or ion implantation techniques all 
of which are kinetically controlled. One of the most sophisticated fabrication tools for this purpose 
is the extreme ultraviolet interference lithography~\cite{solak} where silicon-based two-dimensional 
quantum dot arrays with nanometer size control have been successfully demonstrated.

In this work we tackle the electronic structure and the optical properties of Si NCs along their 
aggregation stages from a computational point of view. For a realistic account of the aggregation 
process, an atomistic electronic structure tool is required that can accommodate about tens of 
thousand nanocrystal and embedding lattice atoms with very irregular shapes. However, currently 
highly sophisticated \textit{ab initio} density functional theory-based plane wave approaches can 
only tackle systems below thousand atoms even with the help of ultrasoft 
pseudopotentials~\cite{martin-book}. For this purpose, a computationally-efficient pseudopotential-based 
electronic structure tool is developed that can handle realistic nanostructures based on the expansion 
of the wavefunction of the aggregate in terms of bulk Bloch bands of the constituent semiconductors. 
With this tool, the evolution of the electronic states as well as the absorption spectra and in 
particular the oscillator strengths of transitions of interest over their aggregation stages are traced.

\section{Theoretical Details}
We employ a computationally-efficient pseudopotential-based electronic structure tool that can tackle 
on the order of tens of thousand atom nanostructures based on the so-called linear combination bulk 
bands recipe of Wang and Zunger~\cite{wang97,wang99}. The crux of the technique is to expand the wavefunction 
of the NC in terms of bulk Bloch bands of the constituent semiconductors;
\begin{equation}
\psi(\vec{r})=\frac{1}{\sqrt{N}}\sum_{n,\vec{k},\sigma} C^{\sigma}_{n,\vec{k}}\,
e^{i\vec{k}\cdot\vec{r}} u^{\sigma}_{n,\vec{k}}(\vec{r})  \, ,
\end{equation}
where, $u^{\sigma}_{n,\vec{k}}(\vec{r})$ is the cell-periodic part of the Bloch states, $N$ is 
the number of primitive cells within the computational supercell, 
$C^{\sigma}_{n,\vec{k}}$ is the expansion coefficient set to be determined 
and $\sigma$ is the constituent bulk material label which points to either the NC core or the 
embedding media. The atomistic NC Hamiltonian is given by
\begin{equation}
\hat{H}=-\frac{\hbar^2\nabla^2}{2m}+
\sum_{\sigma,\vec{R}_j,\alpha} W^{\sigma}_{\alpha}(\vec{R}_j)\,
\upsilon^{\sigma}_{\alpha}\left( \vec{r}-\vec{R}_j-\vec{d}^{\sigma}_{\alpha}\right) \, ,
\end{equation}
where $m$ is the free-electron mass, $W^{\sigma}_{\alpha}(\vec{R}_j)$ 
is the weight function that usually takes values 
between 0 to 1 depending on the type of atom at the position $\vec{R}_j+\vec{d}^{\sigma}_{\alpha}$ 
(which becomes instrumental especially at the interfaces) and 
$\upsilon^{\sigma}_{\alpha}$ is the screened spherical pseudopotential 
of atom $\alpha$ of the material $\sigma$. We use semiempirical pseudopotentials for Si and Ge 
taken respectively from Refs.~\cite{wang94} and \cite{saravia}. This is one other advantage of 
this simpler approach over the more accurate \textit{ab initio} techniques which run into 
well-known band gap shrinkage due to local density approximation~\cite{martin-book} restoration 
of which is not straightforward.

\begin{figure}[ht!]
\includegraphics[width=9cm]{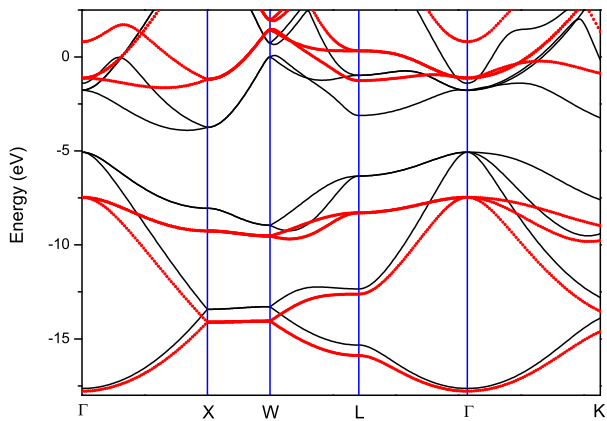}
\caption{\label{bands}Empirical pseudopotential band structure of the Si (solid) and the wide
band gap matrix (circles).}
\end{figure}

An important issue is the choice of the host matrix material. If the NC is surrounded by vacuum, 
this corresponds to the free-standing case. However, the dangling bonds of the surface Si atoms 
lead to quite a large number of interface states which adversely contaminate the effective 
band gap region of the NC. In reality Si NCs are always embedded into a wide band gap host 
matrix which is usually silica~\cite{sput,pecvd}. For this reason, we embed the Si 
NCs into a wide band gap medium having proper band alignment so that both electrons and holes of 
Si will experience quantum confinement. This is illustrated in Fig.~\ref{bands} showing the bulk
band structure of Si and the wide band gap medium with a band gap above 6~eV.
\\

With the possession of the electronic wavefunctions of the NCs, their linear optical properties 
can be readily computed. Within the independent-particle approximation and the artificial 
supercell framework~\cite{weissker} the imaginary part of linear dielectric function becomes 
\begin{eqnarray}
\label{imeps}
\mbox{Im}\/\epsilon_{ii}(\omega) & = & \frac{\left(2\pi e\hbar\right)^2}{mV}\frac{1}{N_k}
\sum_{\vec{k}}\sum_{c,v}\frac{f^{ii}_{cv}(\vec{k})}{E_c(\vec{k})-E_v(\vec{k})}\, \nonumber \\
 & & \times \delta \left({E_c(\vec{k})-E_v(\vec{k})-\hbar\omega} \right)\, ,
\end{eqnarray}
where, $i=x,y,z$ denotes the cartesian components of the dielectric tensor and
\begin{equation}
f^{ii}_{cv}(\vec{k})=\frac{2m\left|\langle c\vec{k}\left| \frac{p_i}{m}\right|v\vec{k}\rangle \right|^2}
{E_c(\vec{k})-E_v(\vec{k})}\, ,
\end{equation}
is the oscillator strength of the transition. In these expressions $V$ is the volume 
of the supercell and $N_k$ is the number of $k$ points chosen within the Brillouin zone,
the label $v$ ($c$) corresponds to occupied (empty) valence (conduction) band states. 
Note that $\mbox{Im}\/\{\epsilon_{ii}\}$ implicitly depends on the filling factor 
$f=V_{\mbox{\begin{scriptsize}{NC}\end{scriptsize}}}/V$
which is the ratio of the NC and the supercell volumes. To obtain a more general quantity 
an effective dielectric function can be introduced as 
$\mbox{Im}\/\{\epsilon_{ii}\}/f$~\cite{weissker2}. 
This is the form we prefer to present our results in the next section; for any specific 
filling factor one can readily infer the corresponding dielectric function. As another 
technical detail a Lorentzian broadening of 0.2~eV of the Dirac delta function 
in Eq.~(\ref{imeps}) is used.
Finally, the absorption coefficient $\alpha(\omega)$ is related to the imaginary part of the 
dielectric function through~\cite{chuang} 
\begin{equation}
\mbox{Im}\/\epsilon(\omega)=\frac{n_r c\epsilon_0}{\omega}\alpha(\hbar\omega)\, .
\end{equation}
This expression is in SI units (unlike the previous expressions) where $n_r$ is the 
index of refraction, $c$ is the speed of light and $\epsilon_0$ is the permittivity of free 
space.

\begin{figure}[ht!]
\includegraphics[width=9cm]{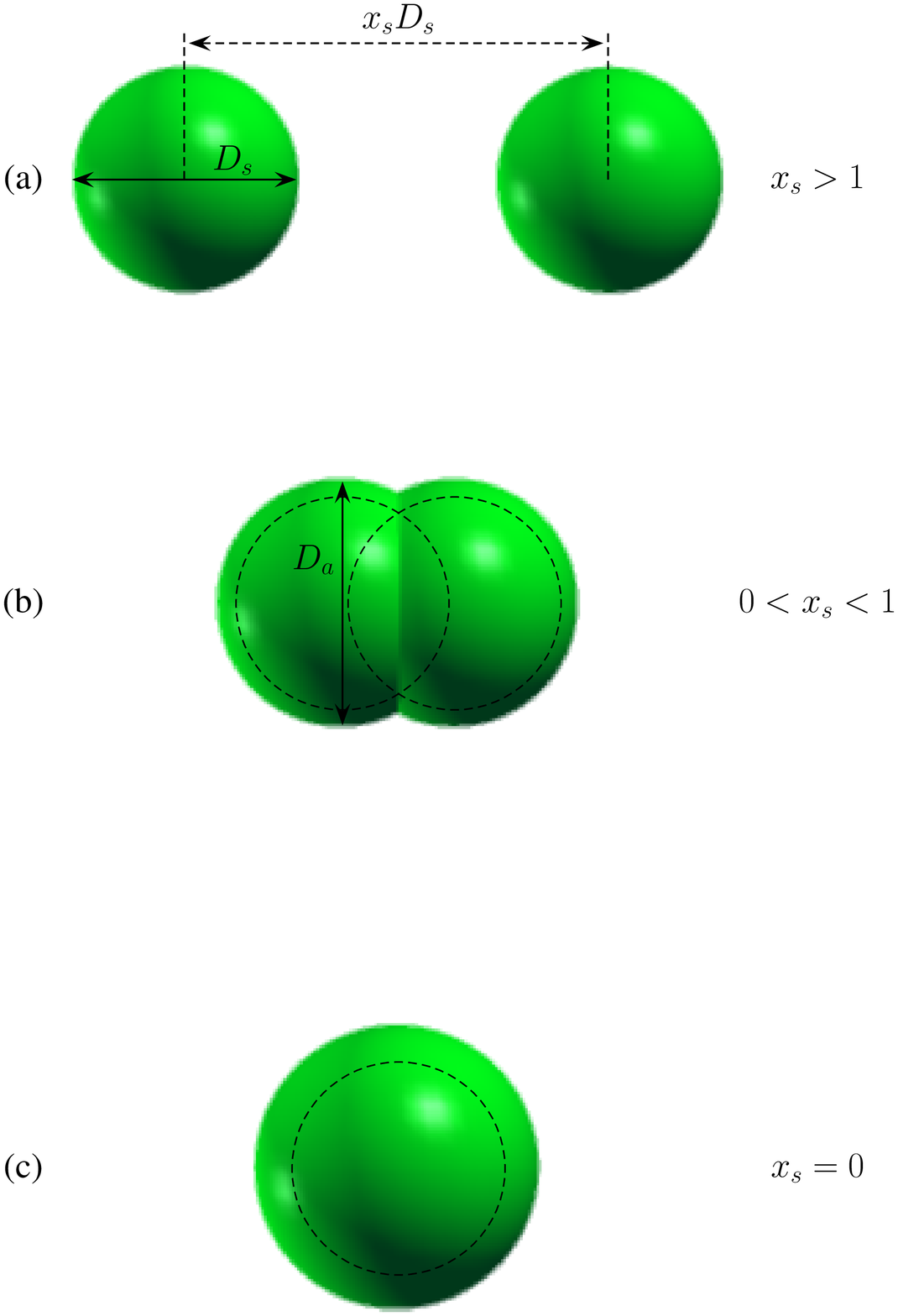}
\caption{\label{aggregation}Illustration of the aggregation stages of two equal-sized NCs}
\end{figure}

\begin{figure}[ht!]
\includegraphics[width=9cm]{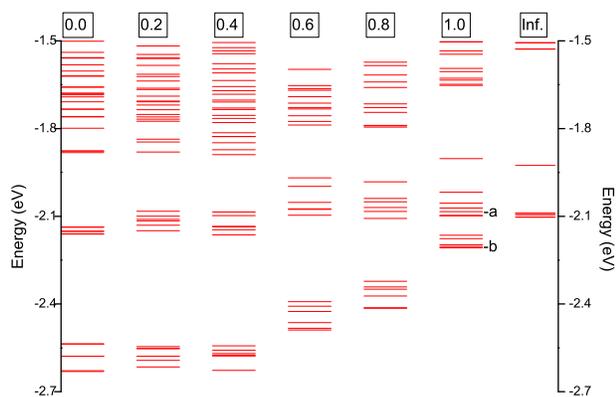}
\caption{\label{CB}Evolution of the low-lying conduction NC states; the $x_s$ values are indicated 
above the spectra. The two states labelled as -a and -b are considered in Fig.~\ref{binding}.}
\end{figure}

\begin{figure}[ht!]
\includegraphics[width=9cm]{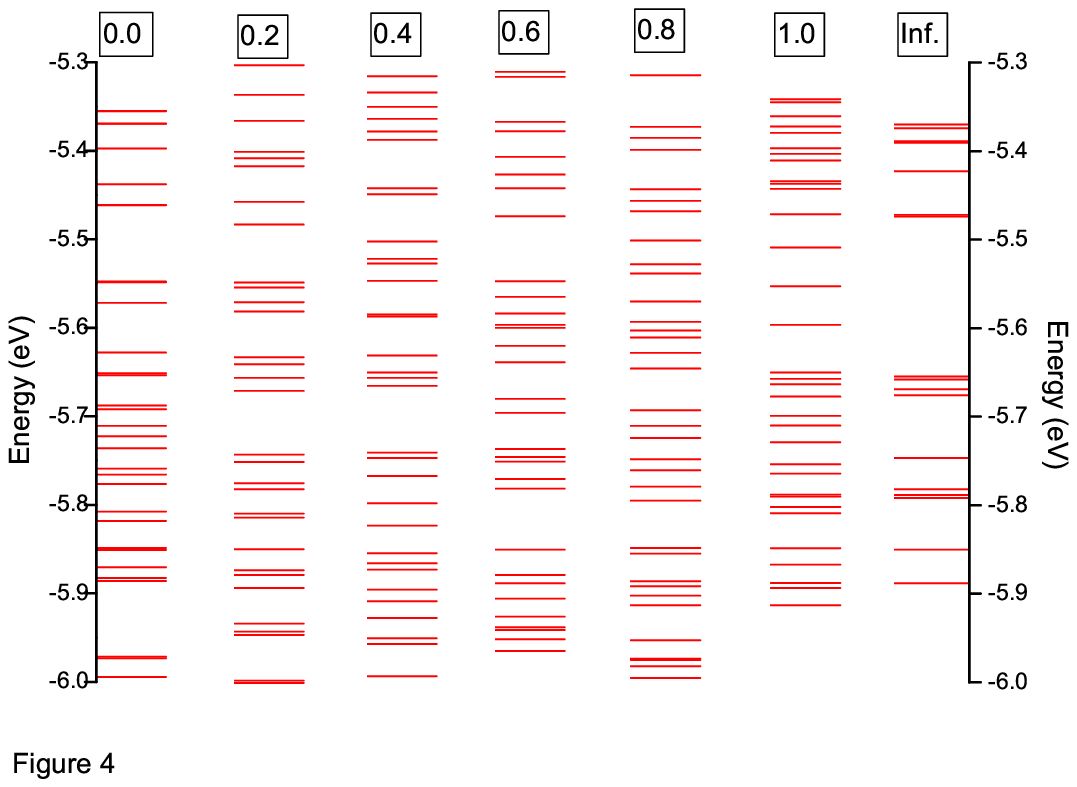}
\caption{\label{VB}Evolution of the low-lying valence NC states; the $x_s$ values are indicated 
above the spectra. }
\end{figure}

Even though there is no geometrical restriction of the approach, we consider the
aggregation of two equal-diameter spherical NCs as illustrated in Fig.~\ref{aggregation}. Here, the
isolated diameters are taken as $D_s$ and the center-to-center distance between the two 
NCs as $x_sD_s$. Assuming no strain generation 
during the aggregation process $(0\le x_s\le 1)$, the diameter of each NC should 
expand from $D_s$ to $D_a$ to preserve the overall volume and the number of atoms. 
A simple geometrical consideration indicates that the expansion factor $y=D_a/D_s$ 
should be determined by solving the following cubic equation
\begin{equation}
y^3+\frac{3x_s}{2}y^2-\left( 2+\frac{x_{s}^{3}}{2} \right)=0\, ,
\end{equation}
for a given $x_s$ value that describes the degree of aggregation. This results in some underestimation
of the number of atoms due to those lost in the reduced surface area of the aggregate; therefore 
we slightly enlarge the aggregate size until we accommodate the correct total number of atoms.

\section{Results}
\begin{figure}[ht!]
\includegraphics[width=9cm]{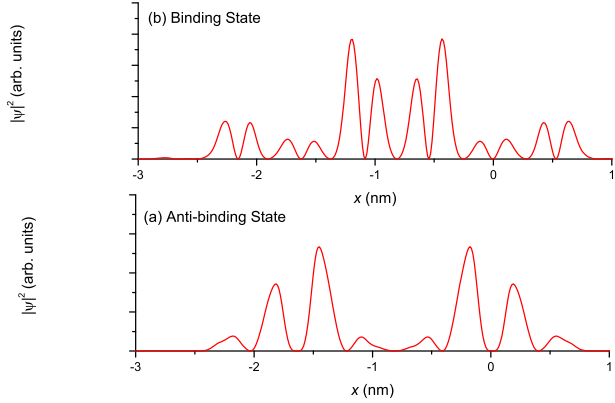}
\caption{\label{binding}Wave function amplitude square plots of the binding and anti-binding 
states indicated in Fig.~\ref{CB} for the $x_s=1$ case.}
\end{figure}

Two NCs each with isolated diameters of 1.6~nm are assumed to approach together along a common $x-$axis. 
The evolution of conduction and valence NC states are shown in Figs.~\ref{CB} and~\ref{VB} respectively,
with the aggregation being quantified by the $x_s$ parameter. In these scales some of the degeneracies
in both the conduction and valence NC states may not be resolved when spherical symmetry is restored 
for $x_s=0$ and $x_s\to \infty$. However, in intermediate aggregation stages these 
degeneracies are further lifted. Also note that the conduction NC states are mainly derived from 
$0.85X$ minima of the bulk Si, hence tend to be six-fold degenerate for the low-lying states. Mixing 
with other bulk band states again lifts the exact degeneracy. From both of these figures, the splitting
of the isolated NC states into binding and anti-binding sets can be observed. This is illustrated in 
Fig.~\ref{binding} where the wave functions of the two states marked as -a (for anti-binding) and -b 
(for binding) are plotted clearly justifying their identification.

\begin{figure}[ht!]
\includegraphics[width=9cm]{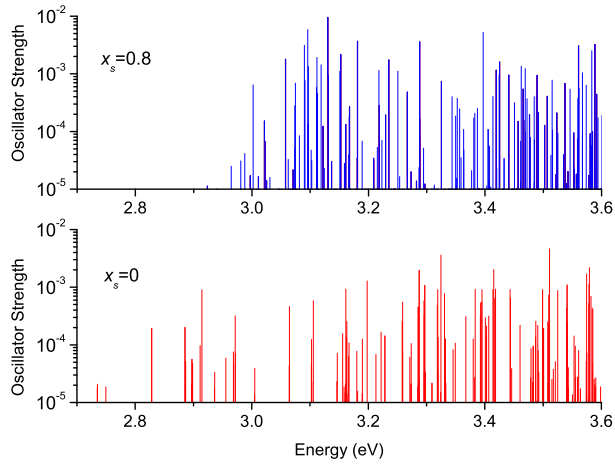}
\caption{\label{osc}The $xx$ component of the oscillator strength tensor of the low-lying transitions for 
$x_s=0.8$ and 0 cases.}
\end{figure}

\begin{figure}[ht!]
\includegraphics[width=9cm]{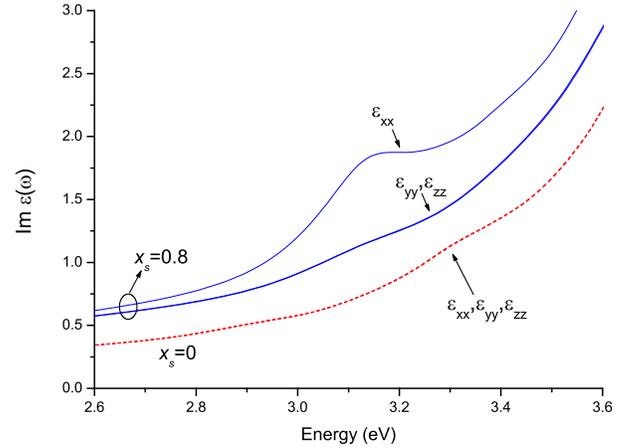}
\caption{\label{eps}The imaginary part of the effective dielectric tensor associated with the previous figure.}
\end{figure}

The effect of NC aggregation on the optical properties is shown in Figs.~\ref{osc} and~\ref{eps} choosing 
 $x_s=0.8$ and 0 cases for comparison. 
 We should remark that the embedding medium has no contribution within the energy window 
 (below 3.6~eV) considered here and all absorption is due to real transitions 
 between NC states over the effective band gap. 
 The oscillator strengths are plotted for the $xx$ component of 
 the tensor where $x$ is the direction along the aggregation. Since the aggregation of the two NCs
 leads to a larger NC, transitions are shifted to lower energies, as expected (cf.~Fig.~\ref{osc}).
 Figure~\ref{eps} demonstrates that $xx$ component of the imaginary part of the dielectric function
 is enhanced whereas the other two components attain a lower but equal value which is also the
 case for the larger and the fully aggregated and spherical $x_s=0$ case. Note that the energy for 
 the local peak of the $xx$ component of the $x_s=0.8$ absorption spectra correlates with strong oscillator 
 strengths in Fig.~\ref{osc} around 3.15~eV which are absent for $x_s=0$. 
 Obviously, in the case of  random growth processes the macroscopic average will wash out such 
 polarization effects, however for more controlled fabrication techniques may reveal these features.
\\
\\
\\
\\

\section{Conclusions}
A pseudopotential-based atomistic electronic structure tool is developed to realistically investigate the 
aggregation process of NCs. With this tool, 
the evolution of the electronic states as well as the absorption spectra and in particular the 
oscillator strengths of transitions of interest over their aggregation stages are traced. 
The predominantly $X$-derived six-pack groupings of conduction NC states are observed.
The low-lying aggregate NC states are identified as the binding and anti-binding splittings
of the isolated states. The effect of the elongation of the NCs in the initial phases of a 
controlled aggregation is reflected as a polarization-dependent absorption spectra. 
Such information may become instrumental with the maturity of the controlled fabrication 
of these NC aggregates.

\begin{acknowledgments}
This work has been supported by the European FP6 Project SEMINANO with the 
contract number NMP4 CT2004 505285 and by the Turkish Scientific and Technical Council 
T\"UB\.ITAK within COST 288 Action.
\end{acknowledgments}

\end{document}